\documentclass[rnote]{aa} 

\usepackage[utf8]{inputenc}

\usepackage{booktabs}
\usepackage{threeparttable}
\usepackage{graphicx}
\graphicspath{{Figures/}}

\usepackage{color}
\usepackage{amsmath}
\usepackage{epstopdf}
\usepackage{url}

\usepackage{xcolor}
\definecolor{green}{HTML}{33CC33}
\definecolor{red}{HTML}{FF3300}
\definecolor{blue}{HTML}{3333FF}

\usepackage[format=plain,labelformat=default, labelsep=endash, labelfont=bf, margin=0pt, tableposition=top, font={footnotesize}]{caption}
\usepackage{subcaption}
\usepackage{times}
\usepackage[english]{babel}
\usepackage{mathrsfs}
\usepackage{xspace}
\newcommand{\code}[1]{\texttt{#1}} 

\newcommand{\ie}{\ensuremath{\rm i.\,e.}\xspace} 
\newcommand{\eg}{\ensuremath{\rm e.\,g.}\xspace}

\newcommand{\Kepler}{\textit{Kepler}\xspace}
\newcommand{\Kp}{\ensuremath{\rm Kp}\xspace}
\newcommand{\ktpt}{\ensuremath{\rm K2P^2}\xspace}
\newcommand{\Kpt}{\ensuremath{\rm\tilde{ K}p_{1}}\xspace}

\renewcommand{\eqref}[1]{Equation~\ref{#1}}
\newcommand{\fref}[1]{Figure~\ref{#1}}
\newcommand{\sref}[1]{Section~\ref{#1}}

\numberwithin{equation}{section}

\makeatletter
\def\maketag@@@#1{\hbox{\m@th\normalfont\normalsize#1}}
\makeatother

\makeatletter
\newcommand\footnoteref[1]{\protected@xdef\@thefnmark{\ref{#1}}\@footnotemark}
\makeatother

\usepackage[varg]{txfonts}

\usepackage{natbib,twoopt}
\usepackage[breaklinks=true,colorlinks=true,citecolor=blue,urlcolor=blue]{hyperref} 
\bibpunct{(}{)}{;}{a}{}{,} 
\makeatletter
\newcommandtwoopt{\citeads}[3][][]{\href{http://adsabs.harvard.edu/abs/#3}%
{\def\hyper@linkstart##1##2{}%
\let\hyper@linkend\@empty\citealp[#1][#2]{#3}}}
\newcommandtwoopt{\citepads}[3][][]{\href{http://adsabs.harvard.edu/abs/#3}%
{\def\hyper@linkstart##1##2{}%
\let\hyper@linkend\@empty\citep[#1][#2]{#3}}}
\newcommandtwoopt{\citetads}[3][][]{\href{http://adsabs.harvard.edu/abs/#3}%
{\def\hyper@linkstart##1##2{}%
\let\hyper@linkend\@empty\citet[#1][#2]{#3}}}
\newcommandtwoopt{\citeyearads}[3][][]%
{\href{http://adsabs.harvard.edu/abs/#3}
{\def\hyper@linkstart##1##2{}%
\let\hyper@linkend\@empty\citeyear[#1][#2]{#3}}}
\makeatother

\defcitealias{2014MNRAS.445.2698H}{HL14} 
\defcitealias{2015ApJ...806...30L}{L15} 

\renewcommand{\thesubfigure}{\arabic{subfigure}}

\title{K2P$^{\mathsf{2}}$-reduced data from campaigns 0--4 of the K2 Mission} 
\author{Rasmus~Handberg\inst{\ref{inst1}}\thanks{\email{rasmush@phys.au.dk}}
\and
Mikkel~N.~Lund\inst{\ref{inst2},\ref{inst1}}\thanks{\email{lundm@bison.ph.bham.ac.uk }} 
}

\institute{Stellar Astrophysics Centre, Department of Physics and Astronomy, Aarhus University, Ny Munkegade 120, DK-8000 Aarhus C, Denmark\label{inst1}
\and School of Physics and Astronomy, University of Birmingham, Edgbaston, Birmingham, B15 2TT, UK\label{inst2}
}

\date{Received 5. Oct. 2015 / Accepted <date>}

\abstract{After the loss of a second reaction wheel the \Kepler mission was redesigned as the K2 mission, pointing towards the ecliptic and delivering data for new fields approximately every 80 days. The steady flow of data obtained with a reduced pointing stability calls for dedicated pipelines for extracting light curves and correcting these for use in, \eg, asteroseismic analysis.} 
{We provide corrected light curves for the K2 fields observed until now (campaigns 0--4), and provide a comparison with other pipelines for K2 data extraction/correction. } 
{Raw light curves are extracted from K2 pixel data using the ``K2-pixel-photometry'' (\ktpt) pipeline, and corrected using the KASOC filter.} 
{The use of \ktpt allows for the extraction of the order of $90.000$ targets in addition to $70.000$ targets proposed by the community --- for these, other pipelines provide no data. We find that \ktpt in general performs as well as, or better than, other pipelines for the tested metrics of photometric quality. In addition to stars, pixel masks are properly defined using \ktpt for extended objects such as galaxies for which light curves are also extracted.} {}


\begin{document}
\maketitle


\section{Introduction}\label{sec:intro}
During May of 2013 a second of four on-board reaction wheels of the \Kepler spacecraft was lost and with it the ability to maintain 3-axis pointing stability of the spacecraft. This led to the redesigned mission ``K2'' where fields towards the ecliptic are observed for a duration of approximately 80 days \citepads{2014PASP..126..398H}. The specific challenges with data quality, combined with the high and steady flow of data from the K2 mission calls for dedicated data analysis pipelines to deliver data to the community.

We here report on the release of light curves extracted from raw pixel data from K2's Campaigns (C) 0--4 using masks defined by the \ktpt (K2-Pixel-Photometry) pipeline presented in \citetads[][]{2015ApJ...806...30L}, hereafter L15.
Corrections of the resulting raw light curves are made using the KASOC pipeline by \citetads[][]{2014MNRAS.445.2698H}, hereafter HL14.
This paper will serve as a data release note, describing the characteristics of the data and the reduced products that have been made available via the KASOC database\footnote{\url{http://kasoc.phys.au.dk/}}.

Our paper is structured as follows. In Sections~\ref{sec:dataext} and \ref{sec:datproc} we briefly describe the concept of the \ktpt pipeline, and the KASOC pipeline which removes systematic artefacts. Here we also report on the properties of the pipeline segment pertaining to defining masks and estimating magnitudes. \sref{sec:datchar} reports on characteristics of the extracted light curves, focusing on noise properties. We detail in \sref{sec:datprod} the data products that have been made available on KASOC and summarise in \sref{sec:dis} with an outlook on future potential updates to the pipeline.

\section{Light curve extraction (\ktpt)}\label{sec:dataext}
Raw light curves were extracted from K2 pixel data from C0--4 using the \ktpt pipeline \citepalias{2015ApJ...806...30L}. Briefly, \ktpt defines pixel masks from a time-summed image of a given EPIC postage stamp where the background has been corrected for. In the summed image, potential pixels to include in defining the pixel masks are then selected based on a flux threshold; clusters in these pixels, each seen as an individual target in the frame, are then located using an unsupervised clustering algorithm; for each pixel-cluster an image-segmentation algorithm is run to segment clusters containing two or more close targets. From the identified targets we extract the flux together with the flux-weighted centroid as a function of time for subsequent correction of the light curves (see \sref{sec:datproc} below).

We note the following modifications to \ktpt as compared to \citetalias{2015ApJ...806...30L}: (1) From C3 onwards background subtraction has been performed by the K2 Science Team on the raw pixel data, so from C3 this step is omitted in the data extraction; (2) From C3 onwards centroids provided in the pixel data (see \sref{sec:datproc}) are used instead of flux weighted centroids calculated from individual targets. The centroids provided in the target pixel data are computed from a select number of bright (non-saturated) targets in the field-of-view of the given campaign; the centroids for these targets are then interpolated to the positions of other targets. Generally, the centroids calculated in this manner show a lower scatter than the ones calculated from individual targets, particularly for saturated targets.

For recent applications of the \ktpt pipeline we refer to \citetads{2015ApJ...809L...3S}, \citetads{10.1086/683103}, \citetads{2015arXiv151003347K}, Lund et al. (2016ab, submitted), and Miglio et al. (2016, in press). For other K2 data extraction and correction pipelines we refer to \citetads{2014PASP..126..948V}, \citetads{2014arXiv1412.1827V}, \citetads{2015MNRAS.447.2880A,2016MNRAS.459.2408A}, \citetads{2015ApJ...806..215F}, \citetads{2015MNRAS.454.4159H}, \citetads{2015A&A...579A..19A,2016MNRAS.456.2260A}, \citetads{2015arXiv151206162V}, \citetads{2015arXiv151109069B}, and \citetads{2016MNRAS.456.1137L}.

\subsection{Magnitude estimation}
\begin{figure}
\includegraphics[width=\columnwidth]{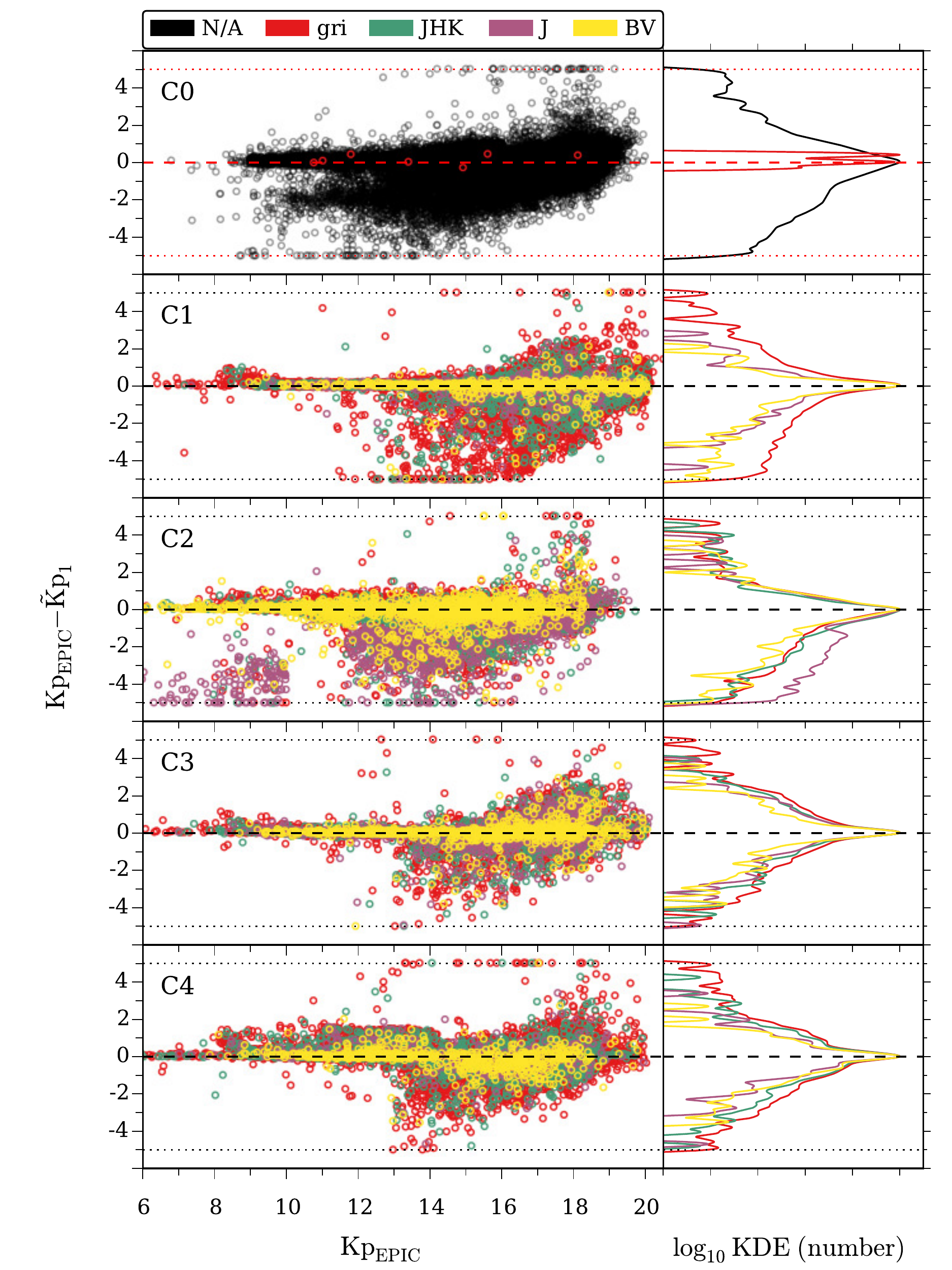}
\caption{Comparison between \Kepler magnitudes from the EPIC ($\rm Kp_{EPIC}$) and the proxy magnitude $\Kpt$ for LC targets. The left part of each panel shows, for different campaigns, the difference in magnitudes against $\rm Kp_{EPIC}$, and colour-coded by the bandpass colours used to compute $\rm Kp_{EPIC}$. The right parts of the panels show KDEs for the differences from each bandpass colour, normalised to the number of targets with that input bandpass colour, and plotted in logarithmic units. Differences have been truncated to the interval between $\pm 5$ (dotted lines). Note that in C0 the $\rm Kp_{EPIC}$ is given by the input colour from the principal investigator of the target, and the information on which colour is used is generally unavailable ($\rm N/A$) in the EPIC.}
\label{fig:mag_vs_mag}
\end{figure}

As described in \citetalias{2015ApJ...806...30L} (see also \citeads{2015MNRAS.447.2880A}) we estimate a proxy for the \Kepler magnitude, $\Kpt$, as
\begin{equation}
	\Kpt = 25.3 - 2.5 \log_{10}(\textit{S}) \, ,
\end{equation}
where $S$ denotes the median of the flux time series extracted for the target (in units of $\rm e^-/s$).
The relation between these magnitudes and the \Kepler magnitude provided in the EPIC, $\rm Kp_{EPIC}$, is shown in \fref{fig:mag_vs_mag} for C0--4 long-cadence (LC; $\Delta t\!\approx\! 29.4\,\rm m$) targets. The different marker colours correspond to the bandpass magnitude(s) used to estimate $\rm Kp_{EPIC}$ following the relations presented in \citetads{2011AJ....142..112B}, \citetads{2012ApJ...746..123H}, and \citet{bib:EPIC}.
We note that in C0 the $\rm Kp_{EPIC}$ was simply given by the input magnitude from the principal investigator proposing a given target, hence these are often simply given by the bandpass magnitude(s) without a proper conversion to the \Kepler bandpass. We use the \code{KepFlag} entry of the EPIC (not defined in C0) to obtain the source for the computed $\rm Kp_{EPIC}$.

Overall, we see a good agreement between $\rm Kp_{EPIC}$ and $\Kpt$ especially for $\rm Kp_{EPIC}\ {\lesssim} 13$. Exceptions to this are seen for smaller groups of targets, typically computed from specific bandpass magnitudes --- we refer to the release notes\footnote{\url{http://keplerscience.arc.nasa.gov/}} for known problems with $\rm Kp_{EPIC}$ magnitudes for different campaigns. We note that targets with masks containing stars not in the EPIC will likely have a positive difference in \fref{fig:mag_vs_mag} as the flux from these targets is naturally combined in $\Kpt$ but unaccounted for in $\rm Kp_{EPIC}$ which only combine the magnitudes from targets found in the EPIC. 

\subsection{Target identification and statistics}\label{sec:stat}
The \ktpt pipeline allows for the extraction of data for all targets in a given frame, not only the targets associated with the EPIC \citep[Ecliptic Plane Input Catalog;][]{bib:EPIC} identifier for the frame. If more than one target is found in a given frame the additional targets will also often have an EPIC identifier and may also be found in a separate frame associated with that EPIC; other times the target has an EPIC ID but has not been proposed for observations. Targets in this latter case would normally be ignored, but can be treated with the \ktpt pipeline.

The problem of identifying the extracted targets is handled by matching all the extracted targets against the EPIC catalog, which itself is a compilation of several other catalogs. Corrections to the world coordinate solutions (see also \sref{sec:wcs}) are calculated using the full EPIC catalog as described in \citetalias{2015ApJ...806...30L} and targets with corrected positions falling within each aperture are assigned to that aperture. All EPIC IDs which falls within the aperture are stored and used to calculate diagnostic information like contamination metrics and the expected brightness, but only the EPIC ID with the brightest Kepler magnitude is in the end assigned to the aperture. 

For the processing of K2 data reported here, a target must fulfil all of the following requirements in order to be finally validated:
\begin{enumerate}[$\circ$] 
	\item A pixel mask must contain a minimum of 8 pixels.
	\item An aperture must be assigned to an EPIC ID.\label{req:epic}
	\item In a given frame, one of the apertures must be assigned to the main EPIC ID.
	\item If an EPIC ID is associated to more than one extracted pixel mask, only the pixel mask with the largest distance to the edge of the given pixel frame is kept.
	\item If the aperture assigned to the main EPIC ID fails any of the above requirements, all extracted apertures from that frame are rejected.
\end{enumerate}
It should be noted that in this release we are depending on the completeness of the EPIC catalogue for the second requirement to be fulfilled. In the future we will be able to provide new targets to the EPIC by matching the measured position and 
\Kpt of a given identified target to catalogues not currently included in the EPIC and feed this information back to the K2 science office.

\begin{figure}
	\includegraphics[width=0.9\columnwidth]{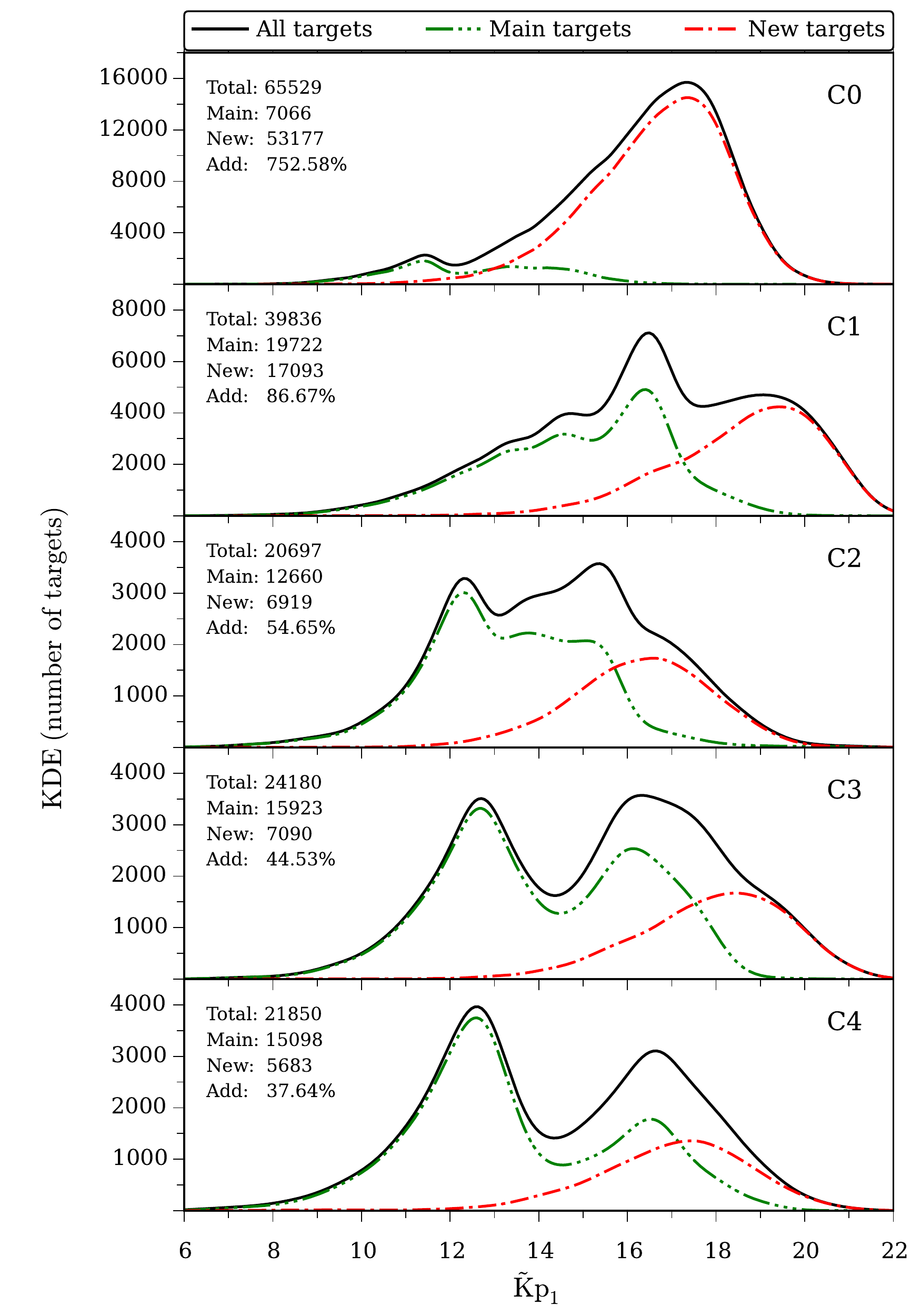}
	\caption{Kernel density estimates (KDE) for the number of targets (with $\Kpt\geq 6$) extracted for each campaign, going from C0 in the top panel to C4 in the bottom panel. The KDEs for each panel denote the total number of targets extracted (full black), the main proposed targets (green dash-dot-dot-dot), and the new targets picked up by \ktpt (red dash-dot). The integrals of the KDEs are listed in each panel together with the percentage addition of targets by \ktpt. Note the change in units on the ordinate for the top panels (C0 and C1).}
	\label{fig:target_stat}
\end{figure}
\fref{fig:target_stat} gives the kernel density estimates (KDEs) for the number of extracted LC targets against $\Kpt$ for the different campaigns. With \ktpt a significant distribution of additional targets (peaking around $\Kpt \sim 18$) is added to the distribution of proposed targets.
In C0 the addition of targets is in excess of $750\%$, while the average addition for following campaigns is around $56\%$. This amounts to an additional ${\sim}90.000$ targets (${\sim}37.000$ not counting C0); the number of proposed targets amounts to ${\sim}70.000$ (${\sim}63.000$ not counting C0).
A clear decreasing trend is seen in the percentage addition of targets with advancing campaigns, because the number of pixels in the so-called ``postage stamps'' around targets has been decreased to optimise the overall pixel budget of the mission.

\subsection{Properties of pixel masks}
\begin{figure}
    \centering
    \includegraphics[width=\columnwidth]{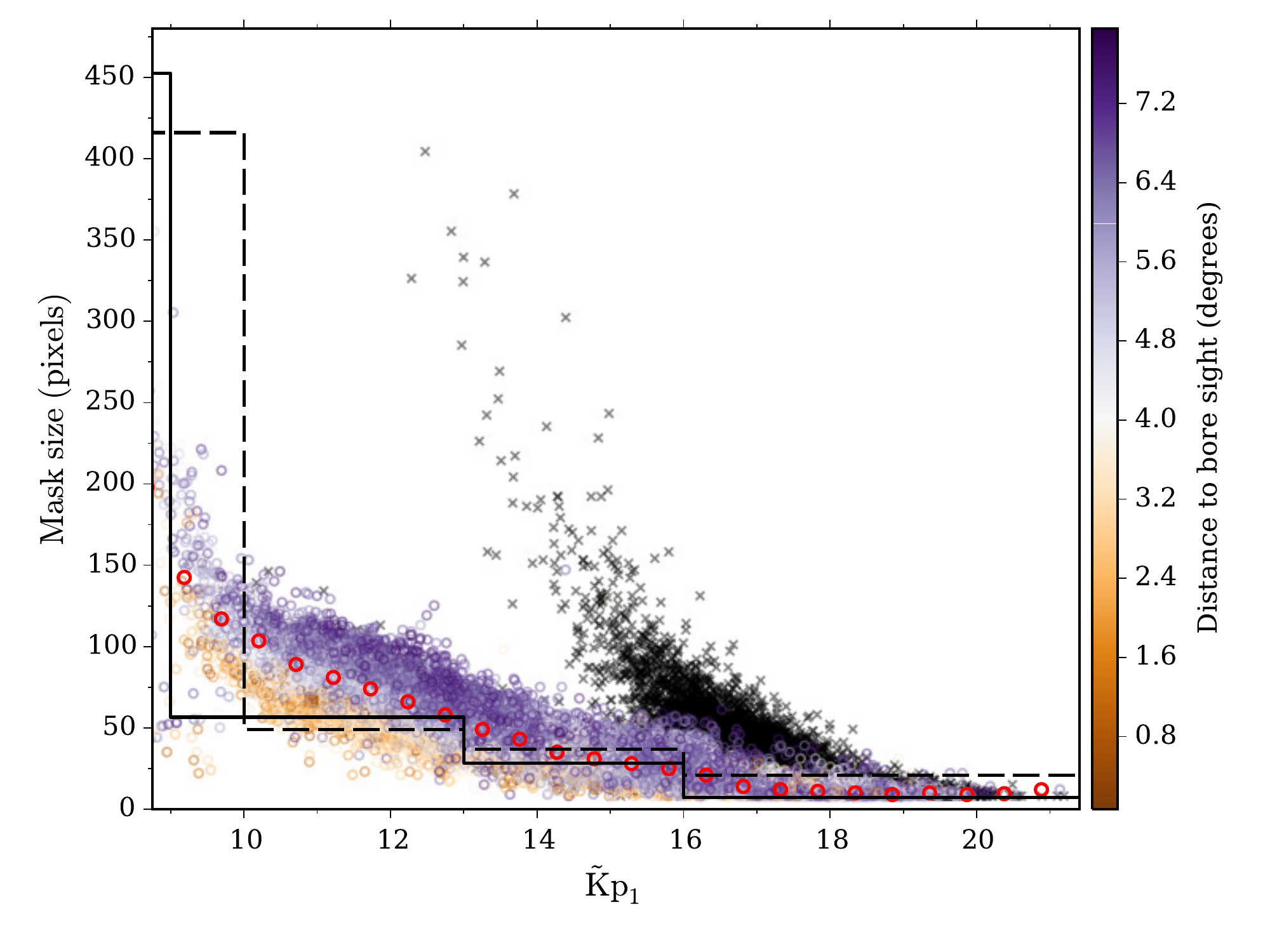}\vspace{-1em}
    \caption{Pixel mask sizes versus the proxy \Kepler magnitude $\Kpt$ for ${\sim}21500$ targets observed in C3, colour-coded with the distance (in degrees) to the spacecraft bore sight; black crosses (``x'') indicate 4143 targets from the GO proposal 3048, which all relate to extragalactic studies of galaxies or active galactic nuclei (AGN), hence these targets are typically truly extended objects not expected to follow the general trend for stars; the red circular markers give the median mask size for $0.5$ wide magnitude bins (not including the extragalactic targets); the full black line shows the mask sizes relation used in \citetads{2015MNRAS.447.2880A} (obtained from engineering data); the dashed black line gives the relation from \citetads{2015A&A...579A..19A}.}
    \label{fig:mag_vs_mask}
\end{figure}

\begin{figure}
\includegraphics[width=\columnwidth]{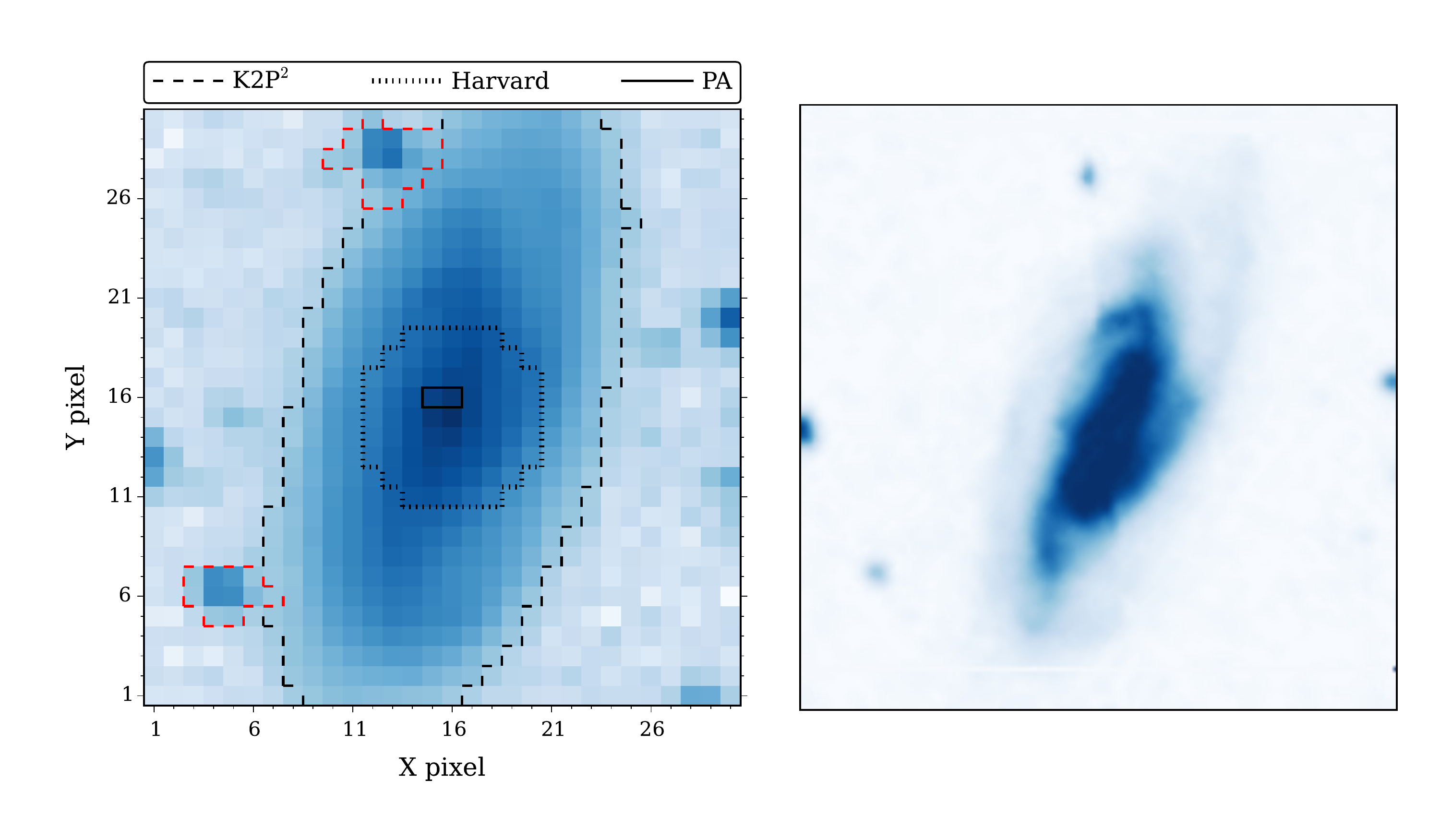}
\caption{Left: Pixel frame for EPIC~206028594, also known as NGC~7300; the colour goes from white (low flux) to blue (high flux). Indicated are the masks defined by \ktpt, the Harvard pipeline \citepads{2014PASP..126..948V}, and the photometric analysis (PA) component of the \Kepler science processing pipeline \citepads{2010ApJ...713L..97B}. We also show (with red masks) the two other targets identified by \ktpt under the constraint of a minimum pixel mask of 8 pixels. Right: Image of NGC 7300 from the Sothern Sky Atlas (SERC), with colours and orientation modified to match in appearance the K2 data.}
\label{fig:galaxy}
\end{figure}

The size of the pixel masks defined by \ktpt is determined by how the flux is distributed for individual targets. One would therefore expect a strong correlation and smoothly varying change in mask size with the stellar magnitude. In \fref{fig:mag_vs_mask} we show the mask sizes as a function of $\Kpt$ obtained for LC targets in C3; as mentioned in \sref{sec:stat} a lower limit on the mask size was set to 8 pixels. A correlation is seen as expected, and as noted in \citetalias{2015ApJ...806...30L} a slight gradient is seen in the mask size with the distance to the spacecraft bore sight from the increase in roll angle.
A tail of dim targets with high magnitudes and large mask sizes is also apparent (marked with black crosses). All of these targets were proposed in the guest observer (GO) proposal 3048 (``The KEGS Transient Survey''), and are all truly extended objects such as galaxies which cannot be expected to follow the same variation as stellar targets. We note that these targets are in the EPIC entry ``Object type'' listed as ``STAR'' (rather than ``EXTENDED''), so care should be exerted when analysing large ensembles of targets from K2 if only the EPIC is used to make the target selection.  
An example of an extended object (here EPIC~206028594 or NGC~7300) is shown in \fref{fig:galaxy} --- as seen, the \ktpt pipeline defines masks for this type of object without any issues or modifications needed. Here we see a disadvantage for this type of target from using the \Kepler photometric analysis (PA) module or the Harvard pipeline \citepads{2014PASP..126..948V} to define masks, which depends on the adopted \Kepler magnitude assuming a stellar target.

\begin{figure*}
    \centering
    \begin{subfigure}[b]{0.33\textwidth}
        \includegraphics[width=\columnwidth]{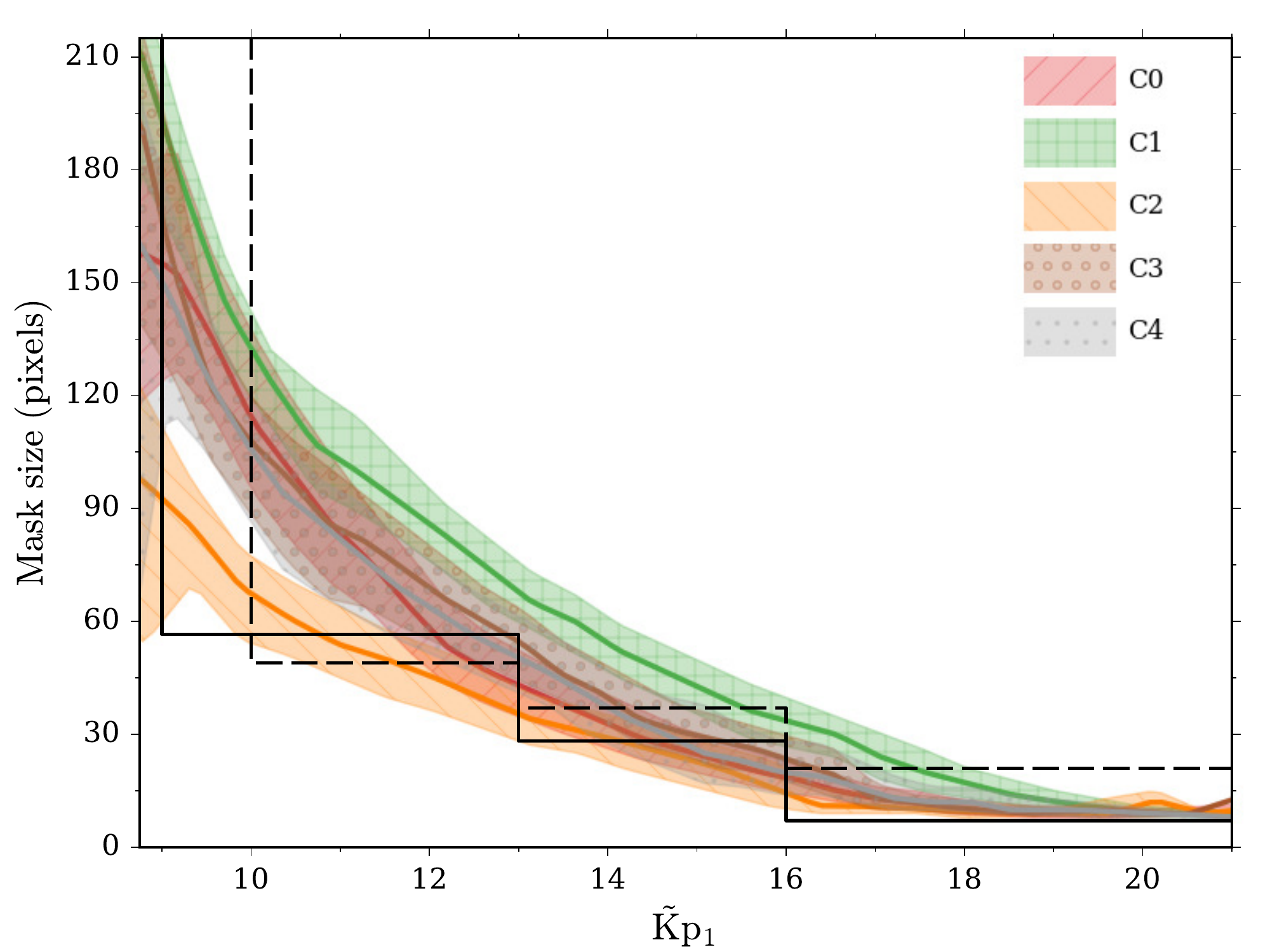}
    \end{subfigure}\hfill
    \begin{subfigure}[b]{0.33\textwidth}
        \includegraphics[width=\columnwidth]{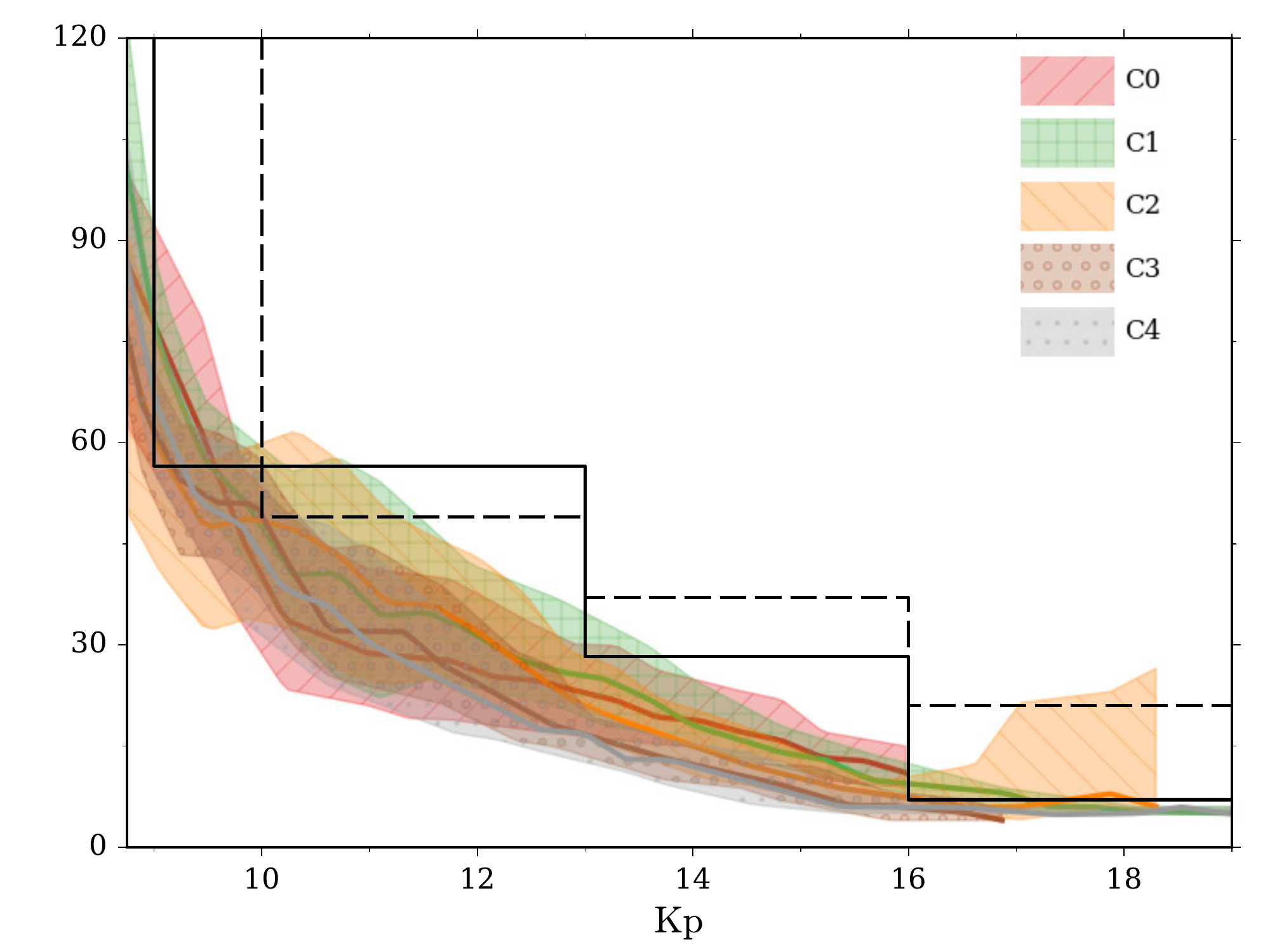}
    \end{subfigure}\hfill
    \begin{subfigure}[b]{0.33\textwidth}
        \includegraphics[width=\columnwidth]{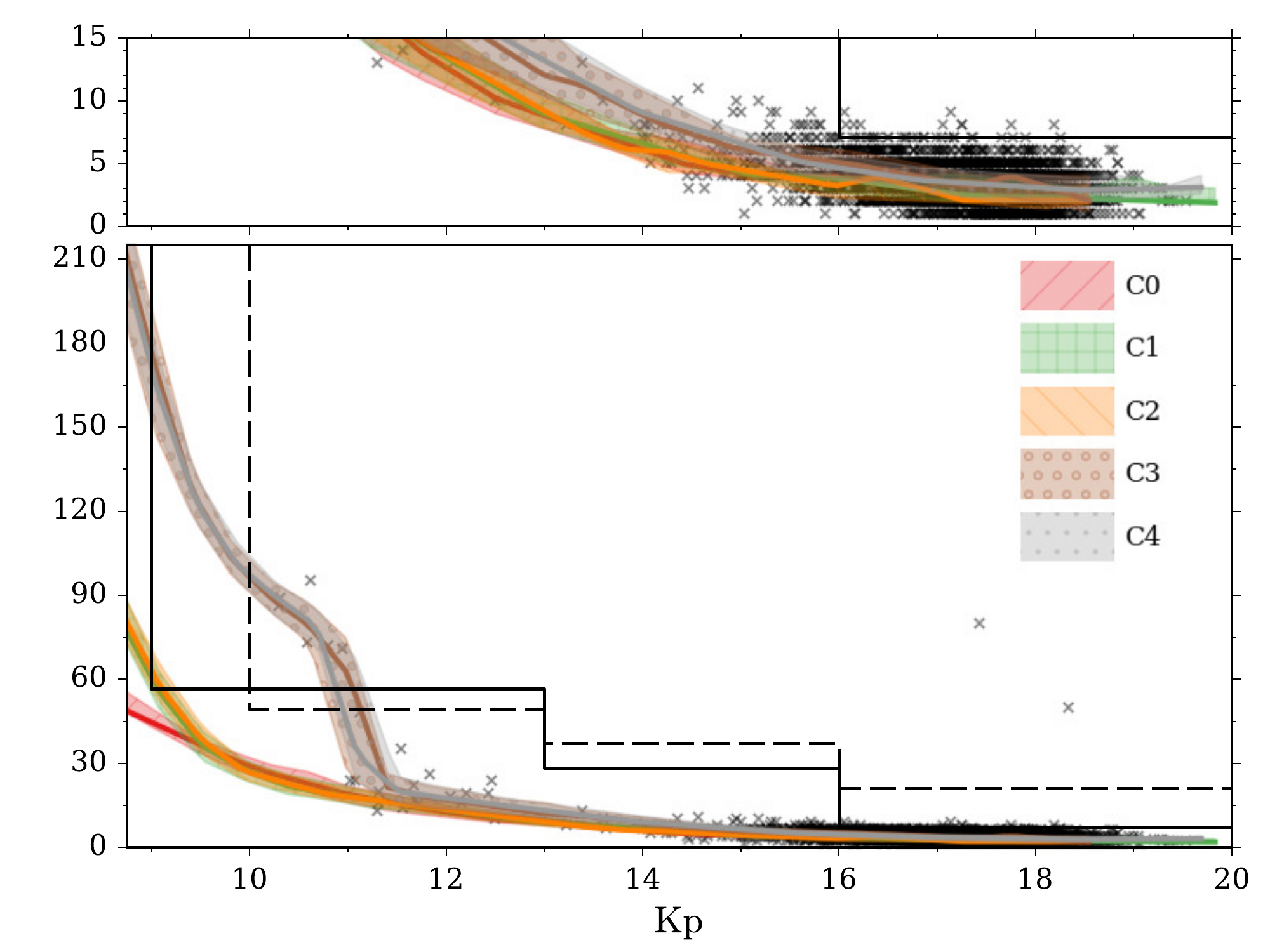}
    \end{subfigure}
    \caption{Comparison between the median binned mask sizes for Campaigns 0--4 excluding targets from AGN/galaxy GO proposals\protect\footnoteref{note1}; hatch-shaded regions indicate the IQR for the mask sizes. For a better visualisation we only show the comparison for $\rm \tilde{K}_{P_{1}}/\ K_{P_{1}}>9$. Left: Mask sizes obtained for \ktpt; shown are also the \citetads{2015MNRAS.447.2880A} (full black) and \citetads{2015A&A...579A..19A} (dashed black) relations, which are given in all panels. Middle: Mask sizes obtained from the \citetads{2014PASP..126..948V} light curves when a pixel-response-function (PRF) fit is performed. Right: Mask sizes from the data products prepared by the photometric analysis (PA) component of the \Kepler science processing pipeline. Black crosses mark the extended targets from the GO proposals 3048. The top panel shows a zoom up to mask sizes of 15 pixels, indicating the possibility of masks down to a single pixel.}
	\label{fig:mag_vs_masks}
\end{figure*}

The panels of \fref{fig:mag_vs_masks} gives the median binned mask sizes --- excluding targets from galaxy/AGN surveys\footnote{\label{note1}we have removed targets from the following GO proposals (see \url{http://keplerscience.arc.nasa.gov/k2-fields.html}): 0009, 0061, 0103, 0106, 1025, 1035, 1072, 1074, 2004, 3004, 3033, 3048, 4038, 4096, 4100.} --- for different campaigns and pipelines, together with the interquartile range (IQR) for the mask sizes. We see for \ktpt a very stable relationship between mask sizes and $\Kpt$ (left panel). We also show the mask size relation obtained by \citetads{2015MNRAS.447.2880A} from engineering data, by \citetads{2015A&A...579A..19A} from C0 (where we note that the authors used \Kp magnitudes without re-calibration), and mask sizes extracted from the reduced light curves by the Harvard pipeline\footnote{obtained from The Mikulski Archive for Space Telescopes (MAST)} (middle panel). From the Harvard pipeline we adopt the mask given as the ``Best'' in their FITS files, but have only included in \fref{fig:mag_vs_masks} the masks defined from pixel-response-function (PRF) fits; the circular masks from the Harvard pipeline for C0 data often span a large range of mask sizes for a given magnitude, with the distinct risk of having multiple targets in a given mask. In general we find that the \citetads{2014PASP..126..948V} masks are smaller than the ones from \ktpt. The right panel shows the mask sizes obtained from the PA component of the \Kepler science processing pipeline. For these we again see masks smaller than those from \ktpt, and notice a change in the definition of mask sizes with magnitude between C2 and C3 where a jump in mask size is seen around $\Kp\approx 11$ from C3 onwards. For both the Harvard and PA masks no noticeable deviation in sizes is seen for the extended objects mentioned above (black crosses in right panel). We also note the allowance in the PA data for masks down to a single pixel, whilst we in \ktpt set a lower limit of 8 pixels.

\subsection{The World Coordinate System}\label{sec:wcs}
\begin{figure}
\includegraphics[width=\columnwidth]{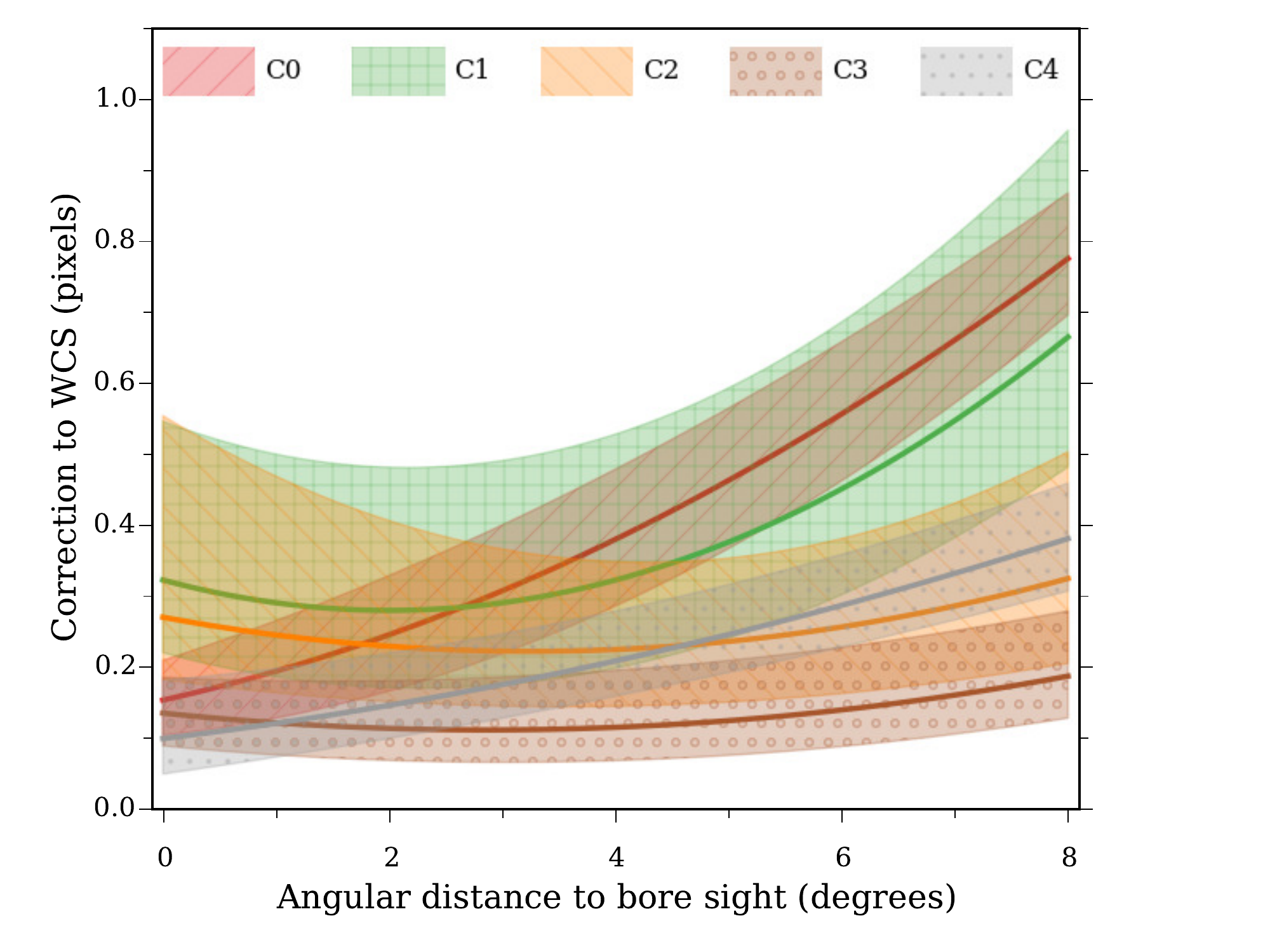}
\caption{Absolute correction in pixels to the estimated position of targets, based on the WCS provided in target pixel files, as a function of angular distance to the spacecraft bore sight. The hatched-shaded region per campaign indicate the IQR of the pixel correction.}
\label{fig:wcs}
\end{figure}

In \citetalias{2015ApJ...806...30L} a correction was introduced to the world-coordinate-system (WCS) provided in the target pixel data.
\fref{fig:wcs} presents the obtained absolute correction in pixels to the expected position of targets from the WCS. We see that the needed correction was the largest in C0, but with later updates to the WCS the correction is seen to drop below $0.4$ pixel. Depending on the crowding of the specific campaign this level of off-set should in general be small enough to correctly identify targets in the frames. As a consequence of this we have changed the maximal shift we allow in the positions to one pixel from C3 and onward.

\section{Light curve corrections (KASOC Filter)}\label{sec:datproc}
The raw light curves are corrected using the KASOC filter \citepalias{2014MNRAS.445.2698H} incorporating a correction for the systematics from the varying roll angle of the spacecraft in a manner mimicking the self-flat-fielding method by \citetads{2014PASP..126..948V}.
The only differences compared to the description of the 1D-correction provided in \citetalias{2015ApJ...806...30L} are (1) the use of centroids from the target pixel files (from C3 onwards) --- these can be found in the \code{POS\_CORR} columns of the target pixel files; (2) the use of the following new ``Quality'' bits: 17 ($\rm decimal\!=\!65536$; ``No data reported'') and 19\footnote{This bit is listed as ``21'' in the pipeline release notes (\url{http://keplerscience.arc.nasa.gov/K2/pipelineReleaseNotes.shtml}) which we believe to be an error.} ($\rm decimal\!=\!262144$; ``Definite Thruster Firing'') for which we exclude the data points. 
We refer to \citetalias{2014MNRAS.445.2698H} for further details on the KASOC filter.

\section{Noise properties}\label{sec:datchar}
\begin{figure*}
 	\centering
	\begin{subfigure}[b]{0.48\textwidth}
        \includegraphics[width=\columnwidth]{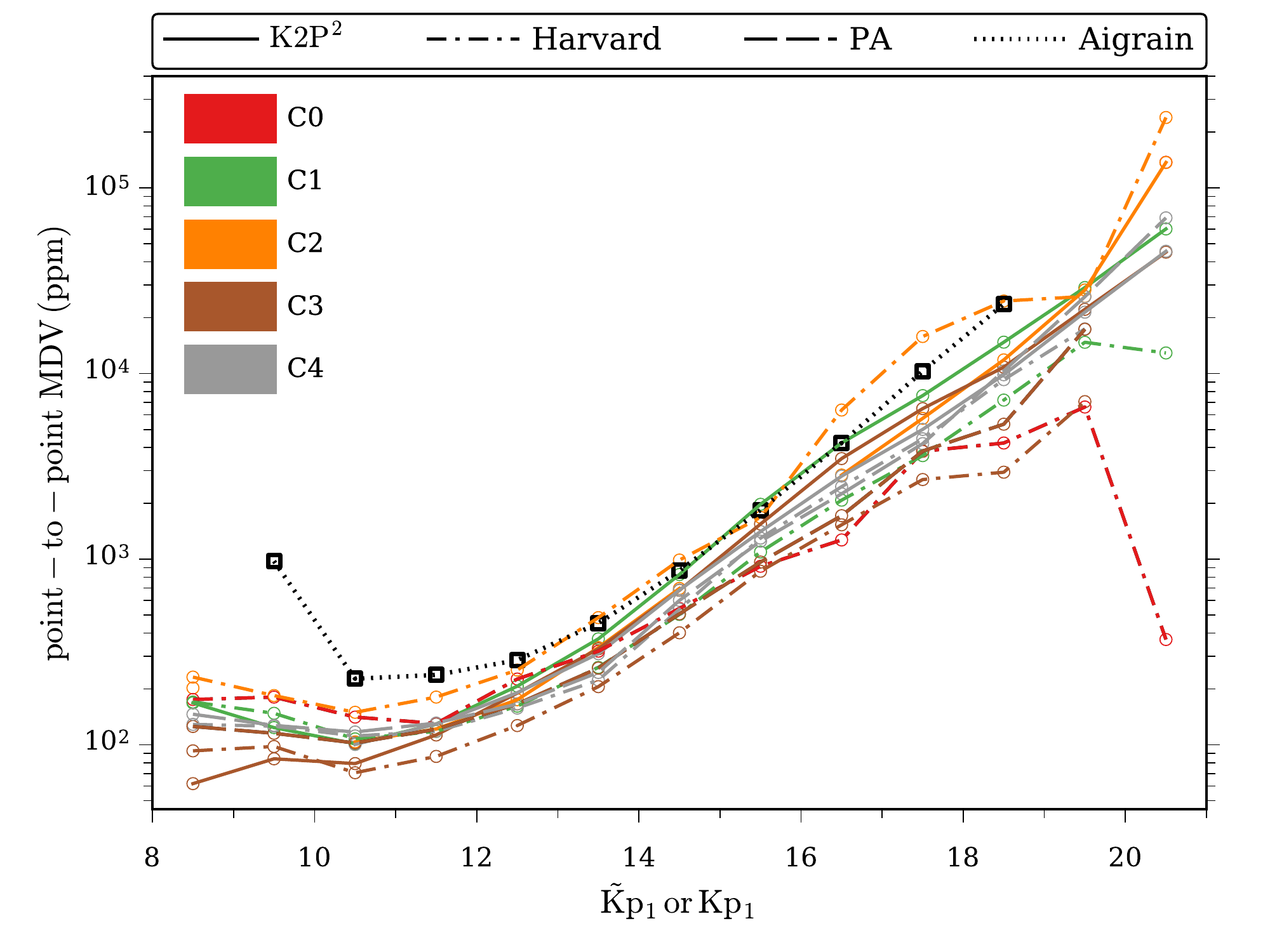}
		\caption{}
		\label{fig:mag_vs_p2p}
	\end{subfigure}~%
	\begin{subfigure}[b]{0.48\textwidth}
		\includegraphics[width=\columnwidth]{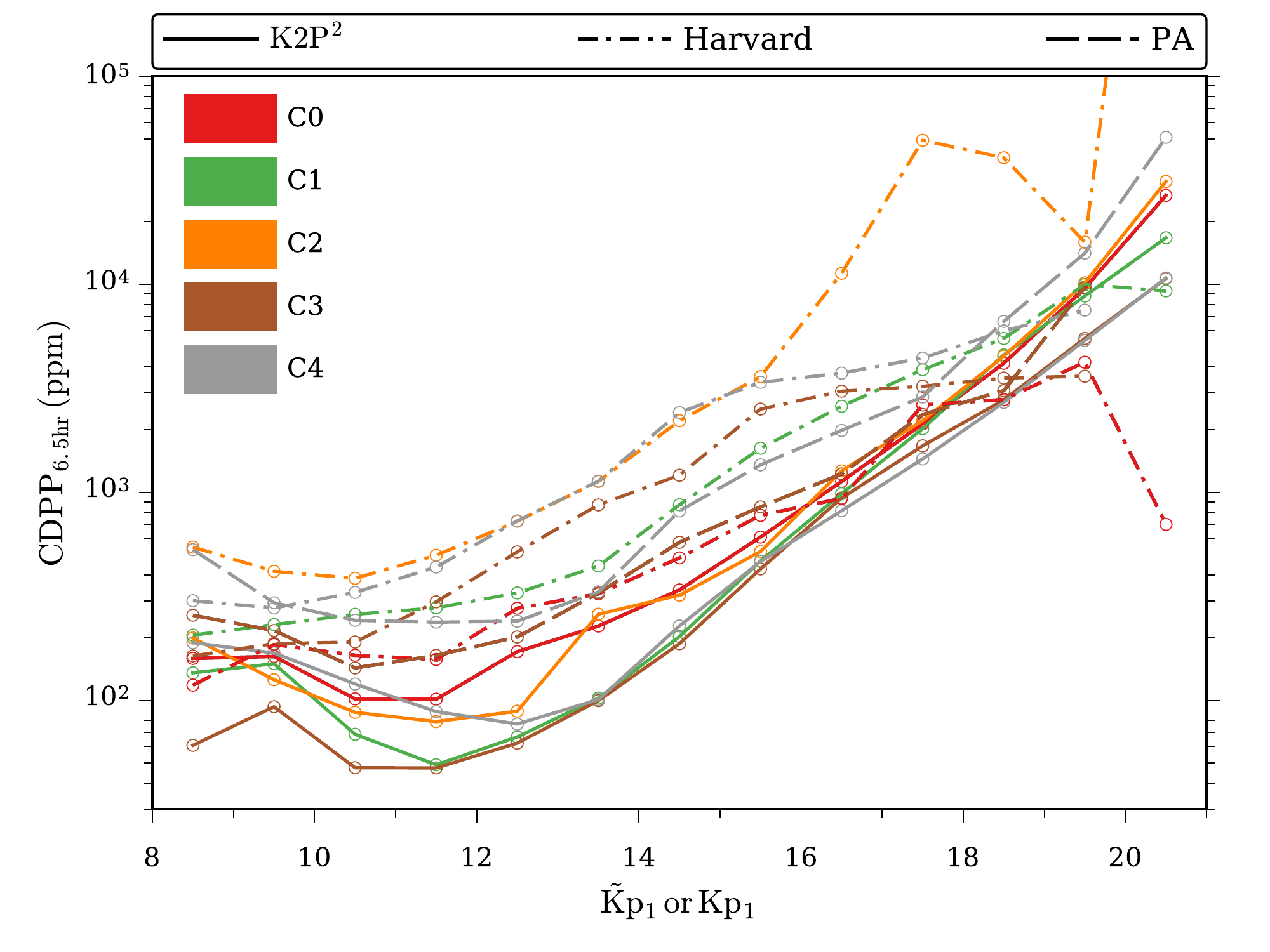}
		\caption{}
         \label{fig:mag_vs_cdpp1}
	\end{subfigure}\\\vspace{0.5em}
	\begin{subfigure}[b]{0.48\textwidth}
        \includegraphics[width=\columnwidth]{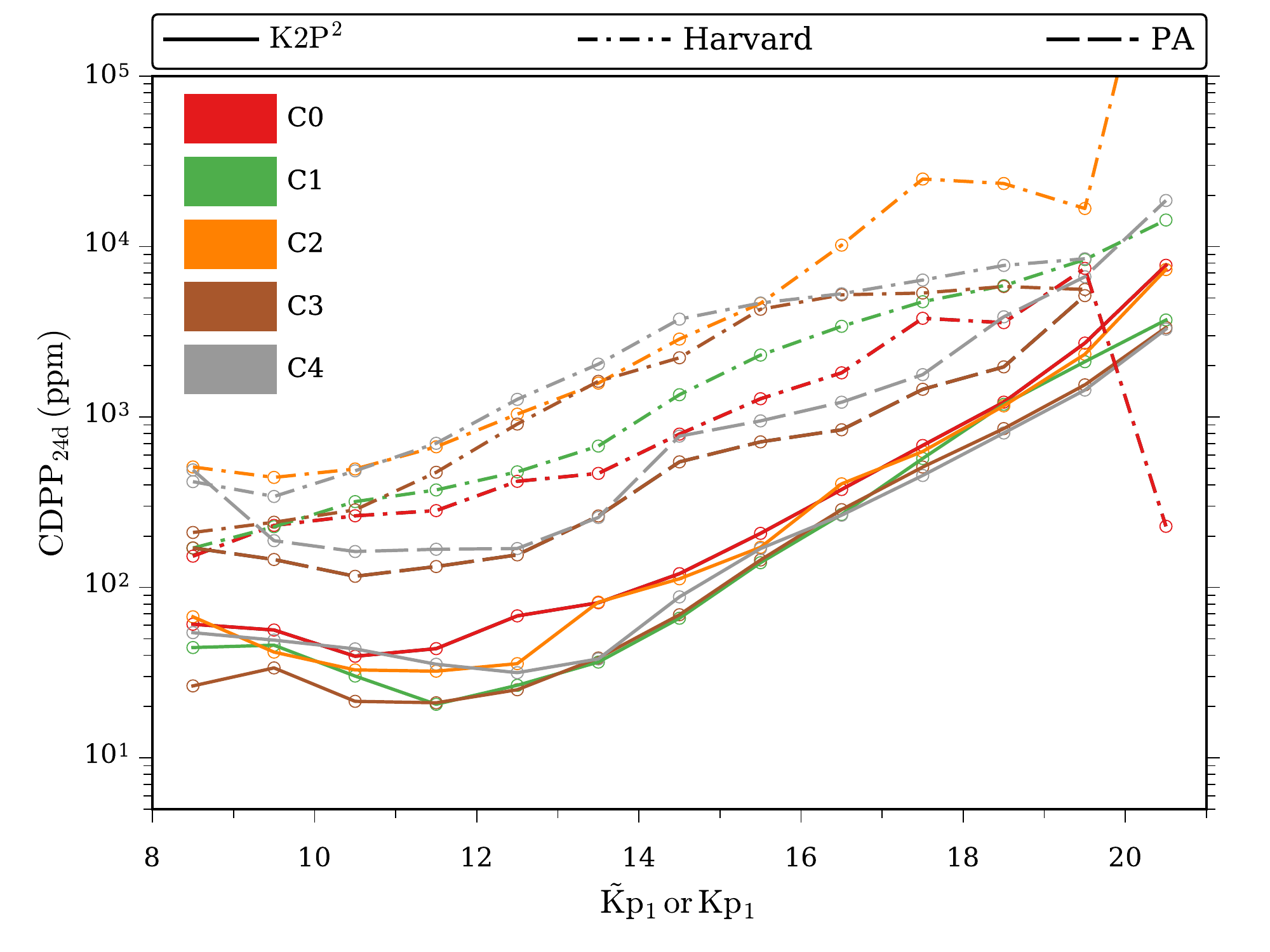}
        \caption{}
        \label{fig:mag_vs_cdpp2}
    \end{subfigure}\qquad%
    \begin{subfigure}[b]{0.48\textwidth}
        \includegraphics[width=\columnwidth]{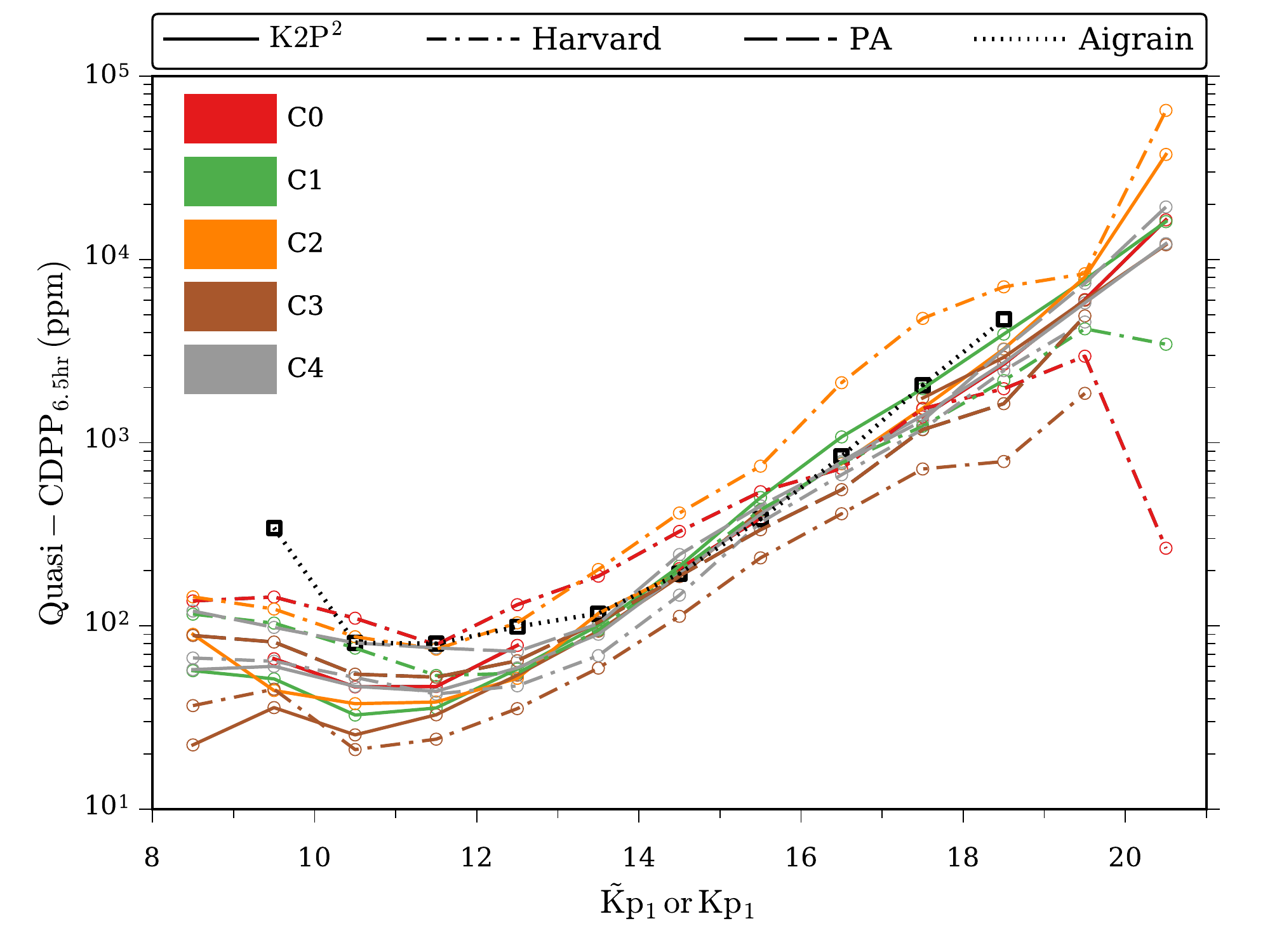}
        \caption{}
        \label{fig:mag_vs_med_pow}
    \end{subfigure}\\\vspace{0.5em}
	\begin{subfigure}[b]{0.48\textwidth}
		\includegraphics[width=\columnwidth]{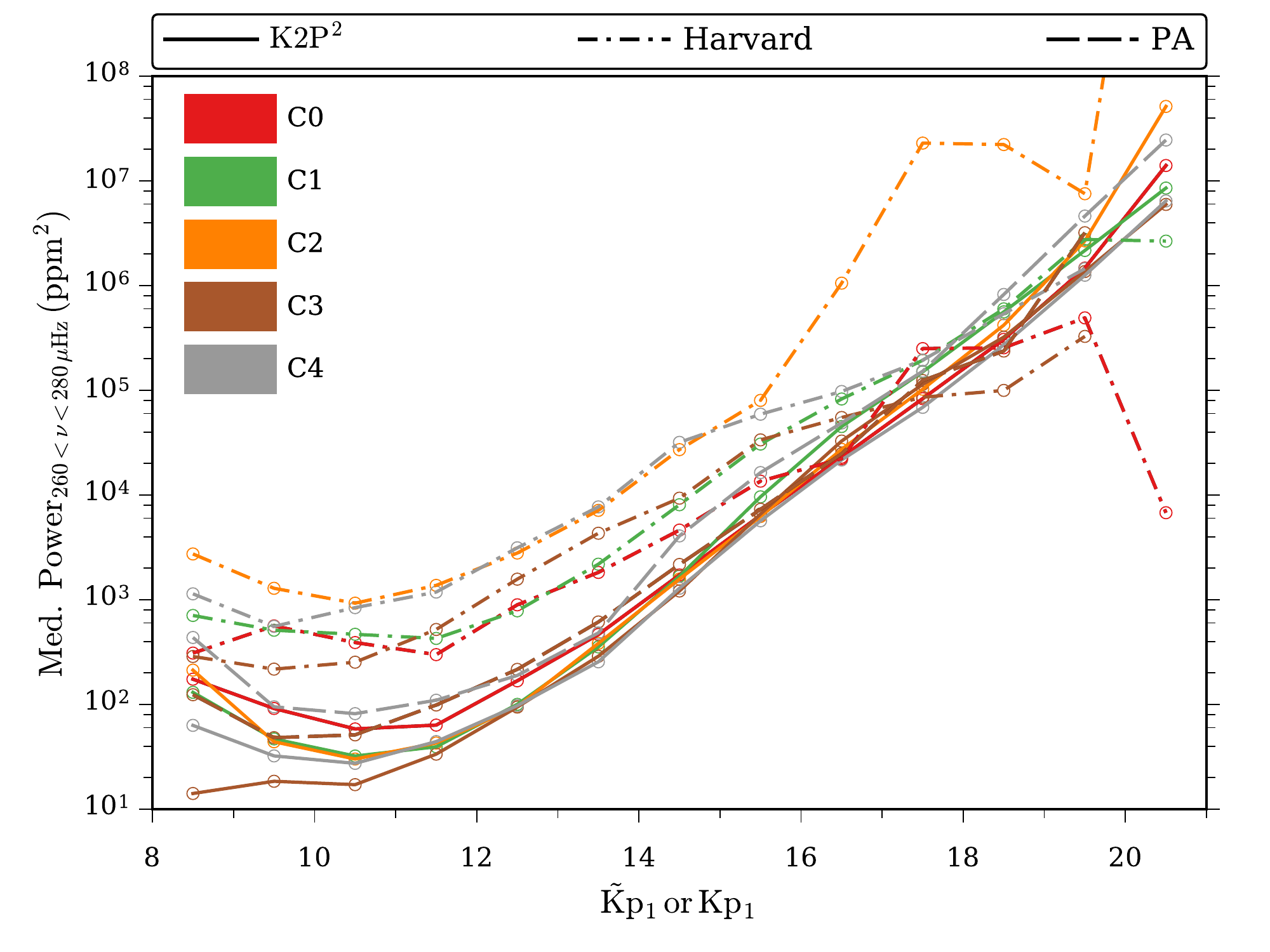}
		\caption{}
		\label{fig:noise_pow}
	\end{subfigure}
    \caption{Comparison of photometric variability metrics between the \ktpt, Harvard (C0--4), and PA (C3--4) pipelines as a function of \Kp (\Kpt for \ktpt). The metrics are depicted by their median magnitude-binned values. (\subref{fig:mag_vs_p2p}): point-to-point median difference variability (MDV). (\subref{fig:mag_vs_cdpp1}): 24-day ``long-CDPP'' (see \citeads{2015AJ....150..133G}). (\subref{fig:mag_vs_cdpp2}): 6.5-hour CDPP, computed following \citetads{2011ApJS..197....6G} (see text). Here we have included the lower envelope for the CDPP$_{\rm 6.5hr}$ values computed in \citetads{2011ApJS..197....6G} from nominal \Kepler data. (\subref{fig:mag_vs_med_pow}): Quasi-CDPP$_{\rm 6.5hr}$ as reported by \citetads{2014PASP..126..948V} and \citetads{2015MNRAS.447.2880A}. For the point-to-point MDV and quasi-CDPP$_{\rm 6.5hr}$ we have included the median binned values obtained by \citetads{2015MNRAS.447.2880A} from engineering data and circular apertures with a 3-pixel radius. (\subref{fig:noise_pow}): Median levels of (unweighted) power spectra from corrected light curves between 260 and 280 $\rm \mu Hz$.}
    \label{fig:noise}
\end{figure*}
Here we present the results obtained for noise characteristics of the filtered data. We consider five frequently used indicators for photometric variability, namely, (1) the point-to-point median difference variability (MDV), given by the median of the time series of point-to-point differences of the corrected light curve \citepads{2011AJ....141...20B}; (2) the 6.5-hour combined differential photometric precision (CDPP$_{\rm 6.5hr}$), obtained in a manner similar to \citetads{2011ApJS..197....6G}, \ie, by applying the combined spectral response of a 6.5-hour Savitzky-Golay (SG) filter \citepads{SGfilter} and a 2-day boxcar filer to the un-weighted power spectrum of the filtered light curve (see also \citeads{2012PASP..124.1279C}) --- following the Parseval-Plancherel theorem \citepads{parseval1806,plancerel1910} the CDPP$_{\rm 6.5hr}$ gives the root-mean-squared scatter of the time series on time scales around 6.5 hours; (3) the 24-day ``long-CDPP'' (CDPP$_{\rm 24d}$), which uses a 24-day SG filter together with a 3.25-day boxcar \citepads{2015AJ....150..133G}. This metric is better suited to capture stellar activity variations by giving a measure for the photometric variability on time scales between 8-15 days; (4) the 6.5-hour ``quasi-CDPP'' metric reported in \citetads{2014PASP..126..948V} and \citetads{2015MNRAS.447.2880A}. This metric is calculated from the median of the rolling 6.5-hour (13 LC cadences) standard deviation divided by $\sqrt{13}$ --- this therefore gives the median of the error on the mean in 6.5-hour window, which is different from the definition of the CDPP by \citetads{2011ApJS..197....6G}. In many ways this metric is close to the point-to-point MDV metric, and is in using the median largely insensitive to any non-stationary components in the corrected light curve, such as residual trends from the ${\sim}6$-hour pointing correction; (5) the median level of the power spectrum from corrected light curves between 260 and 280 $\rm \mu Hz$.

To allow for a comparison with data from the Harvard pipeline and from the \Kepler PA module we have computed the five metrics listed above in a uniform manner for all data sets. From the PA module only C3--4 corrected data have been made available.  
For the computations of metrics 2, 3, and 5 a un-weighted power spectrum was used. In the data that will be made available we will also include a weighted power spectrum, which likely will have lower values for metric 5 (see \sref{sec:datprod}).   

\fref{fig:noise} gives the computed metrics as a function of \Kp (for Harvard and PA) or \Kpt (for \ktpt); the plotted values give the median binned values of the different metrics. For CDPP$_{\rm 6.5hr}$ we include a comparison with the lower envelope of this metric from the nominal \Kepler mission, obtained as the 1st percentile from 0.3 magnitude wide bins of the the values in \citetads{2011ApJS..197....6G} (their Figure~4). 

We see that in general \ktpt returns lower values for the metrics 2, 3, and 5 across the full magnitude range; for metrics 1 and 4 the results from the different pipelines are overall in agreement with each other. The PA results typically lie in-between the Harvard and \ktpt values. Of particular interest to asteroseismic studies we see that for metric 5, the median value of the power spectrum between 260 and 280 $\rm \mu Hz$, drops by a factor of ${\sim}10$ between C0 and C3 for \Kpt$\approx9-11$; this should enable a significant increase in the detectability of oscillations for evolved stars (see \citetads{2015ApJ...809L...3S} for seismic studies of red giants in C1).
We stress that any metric of photometric variability used to assess the ``quality'' of the light curves should be considered in the context of their intended use. The optimal light curve for asteroseismic studies is different from, for instance, that sought for in planetary studies.  
Also, the metrics will be influenced by the definition of pixel masks.

\section{Data products}\label{sec:datprod}
We will make corrected light curves available on the KASOC database, together with both weighted and unweighted power spectra as described in \citetalias{2014MNRAS.445.2698H}.
The formatting of the FITS files will generally follow that described in \citetalias{2014MNRAS.445.2698H}. 
Besides the extensions mentioned in \citetalias{2014MNRAS.445.2698H} the FITS files will have the added extension ``\code{APERTURE}'' which contains an image of the \ktpt pixel mask. We have added the following columns (all in in e$^-$/s) to the binary table belonging to the ``\code{LIGHTCURVE}'' extension in order to expand on the ``\code{FILTER}'' column containing the full KASOC filter: ``\code{XLONG}'', ``\code{XSHORT}'', ``\code{XPOS}'', ``\code{XPHASE}'', and ``\code{FLUX$\_$RAW}''. These contain respectively the $\tau_{\rm long}$ and $\tau_{\rm short}$ filter components, the positional correction to the roll of the spacecraft in K2, the phase-folded component (used if light curve contains transits with known period(s)), and lastly the raw uncorrected light curve. The first four components are constructed such that their sum equals ``\code{FILTER}''.

We have added the entries listed in Table~\ref{tab:primaryheader} to the ``\code{PRIMARY}'' extension holding a header containing information on the star, the data used, and the adopted filter parameters for the given star. See \citetalias{2014MNRAS.445.2698H} for the reminder of the entries supplied, and \citetalias{2015ApJ...806...30L} for explanations to the \ktpt related entries.

\begin{table}
\caption{Special keywords in the first/primary extension of the FITS files. }
\label{tab:primaryheader}
\centering
\begin{tabular}{lp{6cm}}
\toprule
Keyword & Description \\
\midrule
\code{QUARTERS}    & K2 observing campaign in the file. \\
\code{KP{\_}MODE}  & \ktpt mode [\code{Normal}/\code{Custom Mask}]. \\
\code{KP{\_}SUBKG} & Has background been subtracted? [T/F]. \\
\code{KP{\_}THRES} & \ktpt threshold of summed image. \\
\code{KP{\_}MIPIX} & \ktpt minimum pixels in mask. \\
\code{KP{\_}MICLS} & \ktpt minimum pixels for cluster. \\
\code{KP{\_}CLSRA} & \ktpt cluster neighbourhood radius. \\
\code{KP{\_}WS}    & Is watershed segmentation used? [T/F]. \\
\code{KP{\_}WSALG} & Watershed weighting [\code{flux}/\code{dist}]. \\
\code{KP{\_}WSBLR} & Gaussian radius for watershed blur. \\
\code{KP{\_}WSTHR} & \ktpt watershed threshold. \\
\code{KP{\_}WSFOT} & \ktpt watershed footprint (for locating basin minima). \\
\code{KP{\_}EX}    & Use \ktpt treatment of saturated targets (if needed)? [T/F]. \\
\code{KF{\_}MODE}  & Filter operation mode [\code{K2P2}/\code{Default}]. \\
\code{KF{\_}POSS}  & Star positions used in KASOC filter [\code{POSS{\_}CORR}/\code{MOM{\_}CENTR}]. \\
\bottomrule
\end{tabular}
\end{table}

\section{Data policies}
If you use the data that have been produced during this release in your scientific works, we kindly request 
\begin{itemize}
\item[$\circ$] To add the following sentence: ``The \Kepler light curves used in this work has been extracted using the pixel data following the methods described in \citet{2015ApJ...806...30L} and corrected following \citet{2014MNRAS.445.2698H}.''
\item[$\circ$] We request the following sentences be added to the acknowledgements section: ``This research has made use of the KASOC database, operated from the Stellar Astrophysics Centre (SAC) at Aarhus University, Denmark. Funding for the Stellar Astrophysics Centre (SAC) is provided by The Danish National Research Foundation. The research is supported by the ASTERISK project (ASTERoseismic Investigations with SONG and \Kepler) funded by the European Research Council (Grant agreement no.: 267864).''
\item[$\circ$] If extra work was required to produce the data used, like tweaking of filter parameters or similar, we request to be added to the author list of any publications using such specially prepared data.
\end{itemize}

\section{Summary and Outlook}\label{sec:dis}
We have presented characteristics of light curves from K2 campaigns 0--4 extracted using the \ktpt pipeline --- these light curves, and their power spectra, will be made available on the KASOC database.
In terms of the data extraction the \ktpt pipeline performs very well, quantified by an addition of ${\sim}90.000$ extra light curves from untargeted stars for which data would not have been available otherwise. Concerning the definition of pixel masks a correlation as expected is obtained between mask sizes and target brightness, and contrary to other pipelines, masks are properly defined even for extended objects such as galaxies. The use of \ktpt light curves would thus increase the likelihood of detecting signals from supernovas and AGNs in these extragalactic targets. The extracted light curves are corrected using the KASOC Filter pipeline, and we find that these overall have lower photometric variability than those from other pipelines --- this could impact the detectability of, for instance, seismic signals. For \ktpt light curves a median drop in our proxy for white noise (see metric 5 in \sref{sec:datchar}) by a factor of ${\sim}10$ between C0 and C3 for \Kpt$\approx9-11$, which should positively affect the detection of oscillations from red giants.

For future light curve processing we note that the use of house-keeping data from the \Kepler spacecraft could improve light curve corrections, because this would allow for a complete mapping between CCD position and apparent movement on the CCD without the need for computing stellar centroids.

We find that the concept of \ktpt holds great potential for use with the upcoming NASA TESS mission (Transiting Exoplanet Survey Satellite; \citeads{2014SPIE.9143E..20R}). TESS will deliver full frame images of a $24^{\circ}\times 96^{\circ}$ field-of-view (FOV) with a cadence of ${\sim}30$ min for a duration of 27 days per field --- here an automatic and robust definition of pixels masks for targets in the FOV will be needed for the optimal utilisation of TESS data.

\begin{acknowledgements} 
The authors wish to thank the entire \Kepler and K2 teams, without whom these results would not be possible. We are grateful to Ronald L. Gilliland for data provided for the comparison of K2 and nominal \Kepler CDPP, and to Thomas Barclay for answering questions on the K2 data products. `Ta' to Bill Chaplin for giving comments to a draft of this paper.

Funding for the Stellar Astrophysics Centre (SAC) is provided by The Danish National Research Foundation. The research is supported by the ASTERISK project (ASTERoseismic Investigations with SONG and \Kepler) funded by the European Research Council (Grant agreement no.: 267864).

M.N.L. acknowledges the support of The Danish Council for Independent Research | Natural Science (Grant DFF-4181-00415).
M.N.L. was partly funded by the European Community's Seventh Framework Programme (FP7/2007-2013) under grant agreement no. 312844 (SPACEINN), which is gratefully acknowledged.

This research has made use of the following web resources: the \href{http://simbad.u-strasbg.fr/simbad/}{SIMBAD database}, operated at CDS, Strasbourg, France; NASAs  \href{http://adswww.harvard.edu}{Astrophysics Data System Bibliographic Services}; the \href{https://archive.stsci.edu/k2/}{Mikulski Archive for Space Telescopes (MAST)}; \href{http://arxiv.org}{ArXiv}, maintained and operated by the Cornell University Library.
\end{acknowledgements}

\bibliography{MasterBIB1,MasterBIB2}

\begin{thebibliography}{29}
\expandafter\ifx\csname natexlab\endcsname\relax\def\natexlab#1{#1}\fi

\bibitem[{{Aigrain} {et~al.}(2015){Aigrain}, {Hodgkin}, {Irwin}, {Lewis}, \&
  {Roberts}}]{2015MNRAS.447.2880A}
{Aigrain}, S., {Hodgkin}, S.~T., {Irwin}, M.~J., {Lewis}, J.~R., \& {Roberts},
  S.~J. 2015, \mnras, 447, 2880

\bibitem[{{Aigrain} {et~al.}(2016){Aigrain}, {Parviainen}, \&
  {Pope}}]{2016MNRAS.459.2408A}
{Aigrain}, S., {Parviainen}, H., \& {Pope}, B.~J.~S. 2016, \mnras, 459, 2408

\bibitem[{{Armstrong} {et~al.}(2016){Armstrong}, {Kirk}, {Lam}, {McCormac},
  {Osborn}, {Spake}, {Walker}, {Brown}, {Kristiansen}, {Pollacco}, {West}, \&
  {Wheatley}}]{2016MNRAS.456.2260A}
{Armstrong}, D.~J., {Kirk}, J., {Lam}, K.~W.~F., {et~al.} 2016, \mnras, 456,
  2260

\bibitem[{{Armstrong} {et~al.}(2015){Armstrong}, {Kirk}, {Lam}, {McCormac},
  {Walker}, {Brown}, {Osborn}, {Pollacco}, \& {Spake}}]{2015A&A...579A..19A}
{Armstrong}, D.~J., {Kirk}, J., {Lam}, K.~W.~F., {et~al.} 2015, \aap, 579, A19

\bibitem[{{Basri} {et~al.}(2011){Basri}, {Walkowicz}, {Batalha}, {Gilliland},
  {Jenkins}, {Borucki}, {Koch}, {Caldwell}, {Dupree}, {Latham}, {Marcy},
  {Meibom}, \& {Brown}}]{2011AJ....141...20B}
{Basri}, G., {Walkowicz}, L.~M., {Batalha}, N., {et~al.} 2011, \aj, 141, 20

\bibitem[{{Brown} {et~al.}(2011){Brown}, {Latham}, {Everett}, \&
  {Esquerdo}}]{2011AJ....142..112B}
{Brown}, T.~M., {Latham}, D.~W., {Everett}, M.~E., \& {Esquerdo}, G.~A. 2011,
  \aj, 142, 112

\bibitem[{{Bryson} {et~al.}(2010){Bryson}, {Tenenbaum}, {Jenkins},
  {Chandrasekaran}, {Klaus}, {Caldwell}, {Gilliland}, {Haas}, {Dotson}, {Koch},
  \& {Borucki}}]{2010ApJ...713L..97B}
{Bryson}, S.~T., {Tenenbaum}, P., {Jenkins}, J.~M., {et~al.} 2010, \apjl, 713,
  L97

\bibitem[{{Buzasi} {et~al.}(2015){Buzasi}, {Carboneau}, {Hessler}, {Lezcano},
  \& {Preston}}]{2015arXiv151109069B}
{Buzasi}, D.~L., {Carboneau}, L., {Hessler}, C., {Lezcano}, A., \& {Preston},
  H. 2015, ArXiv e-prints [\eprint[arXiv]{1511.09069}]

\bibitem[{{Chaplin} {et~al.}(2015){Chaplin}, {Lund}, {Handberg}, {Basu},
  {Buchhave}, {Campante}, {Davies}, {Huber}, {Latham}, {Latham}, {Serenelli},
  {Antia}, {Appourchaux}, {Ball}, {Benomar}, {Casagrande},
  {Christensen-Dalsgaard}, {Coelho}, {Creevey}, {Elsworth}, {Garc}, {Gaulme},
  {Hekker}, {Kallinger}, {Karoff}, {Kawaler}, {Kjeldsen}, {Lundkvist},
  {Marcadon}, {Mathur}, {Miglio}, {Mosser}, {R}, {Roxburgh}, {Silva Aguirre},
  {Stello}, {Verma}, {White}, {Bedding}, {Barclay}, {Buzasi}, {Deheuvels},
  {Gizon}, {Houdek}, {Howell}, {Salabert}, \& {Soderblom}}]{10.1086/683103}
{Chaplin}, W.~J., {Lund}, M.~N., {Handberg}, R., {et~al.} 2015, \pasp, 127,
  1038

\bibitem[{{Christiansen} {et~al.}(2012){Christiansen}, {Jenkins}, {Caldwell},
  {Burke}, {Tenenbaum}, {Seader}, {Thompson}, {Barclay}, {Clarke}, {Li},
  {Smith}, {Stumpe}, {Twicken}, \& {Van Cleve}}]{2012PASP..124.1279C}
{Christiansen}, J.~L., {Jenkins}, J.~M., {Caldwell}, D.~A., {et~al.} 2012,
  \pasp, 124, 1279

\bibitem[{{Foreman-Mackey} {et~al.}(2015){Foreman-Mackey}, {Montet}, {Hogg},
  {Morton}, {Wang}, \& {Sch{\"o}lkopf}}]{2015ApJ...806..215F}
{Foreman-Mackey}, D., {Montet}, B.~T., {Hogg}, D.~W., {et~al.} 2015, \apj, 806,
  215

\bibitem[{{Gilliland} {et~al.}(2011){Gilliland}, {Chaplin}, {Dunham},
  {Argabright}, {Borucki}, {Basri}, {Bryson}, {Buzasi}, {Caldwell}, {Elsworth},
  {Jenkins}, {Koch}, {Kolodziejczak}, {Miglio}, {van Cleve}, {Walkowicz}, \&
  {Welsh}}]{2011ApJS..197....6G}
{Gilliland}, R.~L., {Chaplin}, W.~J., {Dunham}, E.~W., {et~al.} 2011, \apjs,
  197, 6

\bibitem[{{Gilliland} {et~al.}(2015){Gilliland}, {Chaplin}, {Jenkins},
  {Ramsey}, \& {Smith}}]{2015AJ....150..133G}
{Gilliland}, R.~L., {Chaplin}, W.~J., {Jenkins}, J.~M., {Ramsey}, L.~W., \&
  {Smith}, J.~C. 2015, \aj, 150, 133

\bibitem[{{Handberg} \& {Lund}(2014)}]{2014MNRAS.445.2698H}
{Handberg}, R. \& {Lund}, M.~N. 2014, \mnras, 445, 2698

\bibitem[{{Howell} {et~al.}(2012){Howell}, {Rowe}, {Bryson}, {Quinn}, {Marcy},
  {Isaacson}, {Ciardi}, {Chaplin}, {Metcalfe}, {Monteiro}, {Appourchaux},
  {Basu}, {Creevey}, {Gilliland}, {Quirion}, {Stello}, {Kjeldsen},
  {Christensen-Dalsgaard}, {Elsworth}, {Garc{\'{\i}}a}, {Houdek}, {Karoff},
  {Molenda-{\.Z}akowicz}, {Thompson}, {Verner}, {Torres}, {Fressin}, {Crepp},
  {Adams}, {Dupree}, {Sasselov}, {Dressing}, {Borucki}, {Koch}, {Lissauer},
  {Latham}, {Buchhave}, {Gautier}, {Everett}, {Horch}, {Batalha}, {Dunham},
  {Szkody}, {Silva}, {Mighell}, {Holberg}, {Ballot}, {Bedding}, {Bruntt},
  {Campante}, {Handberg}, {Hekker}, {Huber}, {Mathur}, {Mosser}, {R{\'e}gulo},
  {White}, {Christiansen}, {Middour}, {Haas}, {Hall}, {Jenkins}, {McCaulif},
  {Fanelli}, {Kulesa}, {McCarthy}, \& {Henze}}]{2012ApJ...746..123H}
{Howell}, S.~B., {Rowe}, J.~F., {Bryson}, S.~T., {et~al.} 2012, \apj, 746, 123

\bibitem[{{Howell} {et~al.}(2014){Howell}, {Sobeck}, {Haas}, {Still},
  {Barclay}, {Mullally}, {Troeltzsch}, {Aigrain}, {Bryson}, {Caldwell},
  {Chaplin}, {Cochran}, {Huber}, {Marcy}, {Miglio}, {Najita}, {Smith},
  {Twicken}, \& {Fortney}}]{2014PASP..126..398H}
{Howell}, S.~B., {Sobeck}, C., {Haas}, M., {et~al.} 2014, \pasp, 126, 398

\bibitem[{{Huang} {et~al.}(2015){Huang}, {Penev}, {Hartman}, {Bakos}, {Bhatti},
  {Domsa}, \& {de Val-Borro}}]{2015MNRAS.454.4159H}
{Huang}, C.~X., {Penev}, K., {Hartman}, J.~D., {et~al.} 2015, \mnras, 454, 4159

\bibitem[{{Huber} \& {Bryson}(2015)}]{bib:EPIC}
{Huber}, D. \& {Bryson}, S.~T. 2015, Ecliptic Plane Input Catalog,
  (KSCI-19082-008), \url{https://archive.stsci.edu/k2/epic.pdf}

\bibitem[{{Kurtz} {et~al.}(2015){Kurtz}, {Bowman}, {Ebo}, {Moskalik},
  {Handberg}, \& {Lund}}]{2015arXiv151003347K}
{Kurtz}, D.~W., {Bowman}, D.~M., {Ebo}, S.~J., {et~al.} 2015, ArXiv e-prints
  [\eprint[arXiv]{1510.03347}]

\bibitem[{{Libralato} {et~al.}(2016){Libralato}, {Bedin}, {Nardiello}, \&
  {Piotto}}]{2016MNRAS.456.1137L}
{Libralato}, M., {Bedin}, L.~R., {Nardiello}, D., \& {Piotto}, G. 2016, \mnras,
  456, 1137

\bibitem[{{Lund} {et~al.}(2015){Lund}, {Handberg}, {Davies}, {Chaplin}, \&
  {Jones}}]{2015ApJ...806...30L}
{Lund}, M.~N., {Handberg}, R., {Davies}, G.~R., {Chaplin}, W.~J., \& {Jones},
  C.~D. 2015, \apj, 806, 30

\bibitem[{Parseval~des Ch\^{e}nes(1806)}]{parseval1806}
Parseval~des Ch\^{e}nes, M.-A. 1806, M\'{e}moires pr\'{e}sent\'{e}s \`{a}
  l'Institut des Sciences, Lettres et Arts, par divers savans, et lus dans ses
  assembl\'{e}es. Sciences, math\'{e}matiques et physiques. (Savans
  \'{e}trangers.), 1, 638

\bibitem[{Plancherel(1910)}]{plancerel1910}
Plancherel, M. 1910, Rendiconti del Circolo Matematico di Palermo, 30, 298

\bibitem[{{Ricker} {et~al.}(2014){Ricker}, {Winn}, {Vanderspek}, {Latham},
  {Bakos}, {Bean}, {Berta-Thompson}, {Brown}, {Buchhave}, {Butler}, {Butler},
  {Chaplin}, {Charbonneau}, {Christensen-Dalsgaard}, {Clampin}, {Deming},
  {Doty}, {De Lee}, {Dressing}, {Dunham}, {Endl}, {Fressin}, {Ge}, {Henning},
  {Holman}, {Howard}, {Ida}, {Jenkins}, {Jernigan}, {Johnson}, {Kaltenegger},
  {Kawai}, {Kjeldsen}, {Laughlin}, {Levine}, {Lin}, {Lissauer}, {MacQueen},
  {Marcy}, {McCullough}, {Morton}, {Narita}, {Paegert}, {Palle}, {Pepe},
  {Pepper}, {Quirrenbach}, {Rinehart}, {Sasselov}, {Sato}, {Seager},
  {Sozzetti}, {Stassun}, {Sullivan}, {Szentgyorgyi}, {Torres}, {Udry}, \&
  {Villasenor}}]{2014SPIE.9143E..20R}
{Ricker}, G.~R., {Winn}, J.~N., {Vanderspek}, R., {et~al.} 2014, in Society of
  Photo-Optical Instrumentation Engineers (SPIE) Conference Series, Vol. 9143,
  Society of Photo-Optical Instrumentation Engineers (SPIE) Conference Series,
  20

\bibitem[{Savitzky \& Golay(1964)}]{SGfilter}
Savitzky, A. \& Golay, M. J.~E. 1964, Analytical Chemistry, 36, 1627

\bibitem[{{Stello} {et~al.}(2015){Stello}, {Huber}, {Sharma}, {Johnson},
  {Lund}, {Handberg}, {Buzasi}, {Silva Aguirre}, {Chaplin}, {Miglio},
  {Pinsonneault}, {Basu}, {Bedding}, {Bland-Hawthorn}, {Casagrande}, {Davies},
  {Elsworth}, {Garcia}, {Mathur}, {Di Mauro}, {Mosser}, {Schneider},
  {Serenelli}, \& {Valentini}}]{2015ApJ...809L...3S}
{Stello}, D., {Huber}, D., {Sharma}, S., {et~al.} 2015, \apjl, 809, L3

\bibitem[{{Van Cleve} {et~al.}(2015){Van Cleve}, {Howell}, {Smith}, {Clarke},
  {Thompson}, {Bryson}, {Lund}, {Handberg}, \& {Chaplin}}]{2015arXiv151206162V}
{Van Cleve}, J.~E., {Howell}, S.~B., {Smith}, J.~C., {et~al.} 2015, ArXiv
  e-prints [\eprint[arXiv]{1512.06162}]

\bibitem[{{Vanderburg}(2014)}]{2014arXiv1412.1827V}
{Vanderburg}, A. 2014, ArXiv e-prints [\eprint[arXiv]{1412.1827}]

\bibitem[{{Vanderburg} \& {Johnson}(2014)}]{2014PASP..126..948V}
{Vanderburg}, A. \& {Johnson}, J.~A. 2014, \pasp, 126, 948

\end{thebibliography}

\end{document}